# Magnetization switching by spin-orbit torque in an antiferromagnet/ferromagnet bilayer system

**Shunsuke Fukami[1,2], Chaoliang Zhang[3], Samik DuttaGupta[3], and Hideo Ohno[1,2,3,4]**

[1] Center for Spintronics Integrated Systems, Tohoku University, 2-1-1 Katahira, Aoba, Sendai 980-8577, Japan

[2] Center for Innovative Integrated Electronic Systems, Tohoku University, 468-1 Aramaki Aza Aoba, Aoba, Sendai 980-0845 Japan

[3] Laboratory for Nanoelectronics and Spintronics, Research Institute of Electrical Communication, Tohoku University, 2-1-1 Katahira, Aoba, Sendai 980-8577, Japan

[4] WPI Advanced Institute for Materials Research, Tohoku University, 2-1-1 Katahira, Aoba, Sendai 980 8577, Japan

**Spin-orbit torque (SOT)-induced magnetization switching shows promise for realizing ultrafast and reliable spintronics devices. Bipolar switching of perpendicular magnetization via SOT is achieved under an in-plane magnetic field collinear with an applied current. Typical structures studied so far comprise a nonmagnet/ferromagnet (NM/FM) bilayer, where the spin Hall effect in the NM is responsible for the switching. Here we show that an antiferromagnet/ferromagnet (AFM/FM) bilayer system also exhibits a SOT large enough to**

**switch the magnetization of FM. In this material system, thanks to the exchange-bias effect of the AFM, we observe the switching under no applied field by using an antiferromagnetic PtMn and ferromagnetic Co/Ni multilayer with a perpendicular easy axis. Furthermore, tailoring the stack achieves a memristor-like behaviour where a portion of the reversed magnetization can by controlled in an analogue manner. The AFM/FM system is thus a promising building block for SOT devices as well as providing an attractive pathway towards neuromorphic computing.**

Electrical manipulation of magnetization allows us to realize low-power and high-performance solid-state devices. Recent studies have revealed that an in-plane electric current applied to a heterostructure with large spin-orbit interaction and structural inversion asymmetry gives rise to a torque, so-called SOT, which in turn induces a magnetization switching[1-18]. The SOT-induced switching is useful for three-terminal spintronics device that is capable of high-speed and reliable operation[19]. The origin of SOT is considered to arise from the bulk spin Hall effect (SHE) in the NM and/or the heterostructure interface(s). According to the scenario attributed to the SHE, the in-plane current flowing in the NM generates a pure spin current in the orthogonal direction that is manifested in a Slonczewski-like torque and a field-like torque acting upon the magnetization of FM[5,8,9], where the ratio of the spin current to the charge current is represented by the spin Hall angle. Among the two magnetic configurations presented so far, the perpendicular one has an advantage in terms of fast switching[11,16] and scalability; however, the requirement of a static magnetic field being applied collinear with the current to achieve bipolar switching poses a technological challenge. While recent studies have shown that the need for the field can be eliminated by introducing a lateral structural asymmetry[14,15,17], the present work addresses this issue by employing a new material system.

Up to now, the SOT-induced switching has been shown for the structures with NM/FM bilayers, where Pt[1,3,16], Ta[2,10,14,15], W[4], Hf[17], W/Hf[12], Ir-doped Cu[7], and a topological insulator (Bi,Sb)Te[13] have been used as the NM. Also, the focus of the SHE has so far largely been on 4d or 5d nonmagnetic transition metals and their compounds[20,21]. A recent theoretical work revealed that an AFM exhibits an anomalous Hall effect (AHE)[22], which is generally observed in FMs and has the same origin as the SHE. Following the work, an inverse SHE, in which a pure spin current is converted to a charge current, has been observed experimentally in AFM/FM systems[23,24]. These works promise the existence of a direct SHE in AFM that eventually generates SOT. Note that the AFM can exert an internal effective field on the adjacent FM through the exchange-bias[25-27], which has been a core technology for the read heads of the hard-disk drives. Therefore, by replacing the NM by the AFM, external-field-free switching of perpendicular magnetization may be possible in the AFM/FM bilayer systems through the SOT arising from the direct SHE in the AFM.

Here we study the SOT switching in an AFM/FM bilayer system. We develop stacks consisting of an in-plane exchange-biased ferromagnetic Co/Ni multilayer with a perpendicular easy axis on top of an antiferromagnetic PtMn (Fig. 1a). Evidence of direct SHE and resulting SOT-induced switching under zero magnetic fields are obtained. Furthermore, we find that, in the AFM/FM system, the portion of reversed magnetization can be controlled continuously by adjusting the magnitude of the applied current; in other words, the system has the potential to serve as a memristor[28-31], which is useful for neuromorphic computing[32].

**Results**

**Sample preparation and magnetic properties.** The stacks are deposited on Si wafers by sputtering under zero magnetic fields. The structure of the films is, from the substrate side, Ta(3)/ Pt(4)/ PtMn($t_{PtMn}$)/ [Co(0.3)/ Ni(0.6)]$_2$/ Co(0.3)/ MgO(1.2)/ Ta(2) (in nm), with $t_{PtMn}$ = 2.0, 4.0, 5.0, 6.0, 7.0, 7.5, 8.0, 8.5 nm. After the deposition, to provide the exchange bias, the samples are annealed at 300 °C for two hours under an in-plane magnetic field of 1.2 T. Magnetization hysteresis loops (*m-H* loops) in the *X, Y*, and *Z* directions for the samples with $t_{PtMn}$ = 6.0 and 8.0 nm are shown in Fig. 1b,c, respectively, where the *Z* axis is normal to the film plane, and *X* axis is in the direction in which the magnetic field is applied during the annealing. The hysteresis loops in the *X* and *Y* directions overlap with one another for $t_{PtMn}$ = 6.0 nm, whereas that in the *X* direction is shifted to the −*X* direction by 12.7 mT for $t_{PtMn}$ = 8.0 nm as a result of the exchange bias along the +*X* direction. The magnitude of the shift corresponds to the bias field $H_{bias}$. The details of magnetic and transport properties of the blanket films are described in the Supplementary Information. $H_{bias}$ increases with $t_{PtMn}$ above 7.0 nm, whereas below that it is effectively zero. The perpendicular easy axis is obtained for all the samples studied here, and the effective anisotropy field is almost constant in the range of 120 – 150 mT for all the $t_{PtMn}$.

The deposited films are processed into Hall-bar structures with a 10-μm-wide channel along the *X* axis (Fig. 1d). To evaluate the perpendicular component of the magnetization, Hall resistance $R_H$ due to the AHE is measured with channel current $I_{CH}$ of 1 mA, corresponding to a current density of around $2.4 \times 10^9$ A m$^{-2}$. $R_H$ vs perpendicular magnetic field $H_Z$ ($R_H$-$H_Z$ loops) of the Hall devices and corresponding *m*-$H_Z$ loops of the blanket films are shown in Fig. 1ef, respectively, for $t_{PtMn}$ = 6.0, 7.0, 7.5, 8.0, 8.5 nm. As $t_{PtMn}$ increases, the squareness and remanence of the loops decrease because of the increasing exchange bias. The trends for the $R_H$-$H_Z$ loop and the *m*-$H_Z$ loop are consistent.

**$R_H$-$H_Z$ loops under SOT.** Next we examine the variation of $R_H$-$H_Z$ loop as a function of $I_{CH}$, where the magnitude of SOT is expected to scale with the $I_{CH}$. Figure 2a,b shows the $R_H$-$H_Z$ loops for the devices with $t_{PtMn}$ = 6.0 and 8.0 nm, respectively, obtained at $I_{CH}$ = −32, +1, +32 mA. The loops become narrower with increasing $I_{CH}$ for both devices. Coercive field $H_C$, defined as $H_Z$ at which $R_H$ changes by more than half, is plotted as a function of $I_{CH}$ for both down-to-up and up-to-down reversals in Fig. 2c,d. Striking difference is seen between the two samples. For $t_{PtMn}$ = 6.0 nm (non-biased device), the diagram shows fourfold symmetry, that is, the variation of $H_C$ depends on neither the switching direction nor the sign of $I_{CH}$; on the other hand, for $t_{PtMn}$ = 8.0 nm (exchange-biased device), it shows twofold symmetry, that is, the sign of $I_{CH}$ specifies the preferable switching direction. We see the former (latter) trend for all the devices with $t_{PtMn} \leq 6.0$ nm ($\geq 7.0$ nm). Note here that the increase in device temperature due to the Joule heating under the $I_{CH}$ used is evaluated to be much less than 10 K, which is too small to account for the decrease of $H_C$. Also, the Oersted field generated by $I_{CH}$ of 10 mA is calculated to be around 0.6 mT, which is also too small to cause the observed variation in $H_C$.

The obtained results can be interpreted by considering a current-induced SOT with Slonczewski-like symmetry. As depicted in Fig. 2e,f, the direction of the effective field due to the Slonczewski-like SOT ($H_{SL}$) depends on the magnetization polarity and takes the form of **M** × **Ŷ**, where **M** is the magnetization vector and **Ŷ** is the unit vector along the Y axis[5,8,9]. In consideration of the fact that the magnetization tilts in the +X direction in our exchange-biased devices, the result shown in Fig. 2d is accounted for by an action of a clockwise $H_{SL}$ around Y axis. This indicates that the spin Hall angle of PtMn is positive, which is consistent with the previous study on the inverse SHE in PtMn[24]. We also find that reversing the bias direction for the device with $t_{PtMn} \geq 7.0$ nm shows the opposite tendency: a positive current decreases $H_C$ for down-to-up reversal and a negative current does vice versa (see Supplementary Information). This is also in line with the scenario that the Slonczewski-like SOT and the exchange bias govern the switching. Thus, the obtained $R_H$-$H_Z$ loops with various $I_{CH}$ indicate that the antiferromagnetic PtMn shows direct SHE and that the magnetization reversal is governed by both the consequent SOT and exchange bias. Also, the decrease in $H_C$ with increasing $I_{CH}$ for $t_{PtMn}$ = 6.0 nm indicates that the SOT is not related to the presence of the exchange bias. The effective field is evaluated to be 78±5 mT per $10^{12}$ A m$^{-2}$, which includes the contributions of both the Slonczewski-like torque and field-like torque (see Supplementary Information for the evaluation procedure). This magnitude is comparable to a reported value for Ta/CoFeB/MgO (230 mT)[33], suggesting a comparable magnitudes of the SOT for these systems.

**$R_H$-$I_{CH}$ loops under no magnetic field.** Following the $R_H$-$H_Z$ loops measurement that suggests the possibility of external-field-free SOT switching, we study the response of $R_H$ to $I_{CH}$ under zero fields. The measurement sequence of a basic unit is depicted in Fig. 3a. First, an initialization pulse with the magnitude of $I_{INIT}$ (= ±44 mA) is applied to the channel. Then, an application of 0.5-s-long pulses ($I_{CH}$) with various magnitudes (indicated by blue solid lines) and a measurement of $R_H$ (indicated by red triangles) are repeated. In case of positive (negative) $I_{INIT}$, magnitude of $I_{CH}$ is first increased in the negative (positive) direction up to −(+)$I_{MAX}$, next decreased, and then applied to positive (negative) direction up to +(−)$I_{INIT}$.

Figure 3b-d shows the $R_H$-$I_{CH}$ loops measured under zero fields for $t_{PtMn}$ = 8.0 nm. Regardless of the direction of exchange bias and initialization, change in $R_H$ with $I_{CH}$ is observed as expected, indicating that the perpendicular magnetization is reversed solely by the current-induced torque. The direction of the change in $R_H$ with respect to the sign of $I_{CH}$ and exchange bias is consistent with the scenario deduced from the results shown in Fig. 2. Unexpectedly, moreover, $R_H$ at the halfway point of the loop changes gradually according to $I_{MAX}$, instead of showing binary states. Taking Fig. 3b for example, $R_H$ at the halfway point of the loop for $I_{CH}$ = −24 mA is about the half of the value at more than −40 mA. Such behaviour was never seen in previous works on Pt/Co/AlO$_X$[3] and Ta/CoFeB/MgO[2,10] with a similar device geometry, and thus appears to be inherent in the present PtMn/[Co/Ni] system. In Supplementary Information, we show the $R_H$-$I_{CH}$ loops for other devices with different $t_{PtMn}$. We find that the intermediate states are more likely to appear as $t_{PtMn}$ increases. The reason for the gradual change of $R_H$ and its significance will be discussed later.

**$R_H$-$I_{CH}$ loops under various $H_X$.** To retrieve quantitative information concerning the SOT and exchange-bias in the present system, we next study the $R_H$-$I_{CH}$ properties under various magnetic fields along the X axis ($H_X$). Here we scan the $I_{CH}$ ranging from −35 to +35 mA under a static $\mu_0 H_X$ ranging from −40 to +40 mT. Figure 4a,b shows the obtained $R_H$-$I_{CH}$ loops for $t_{PtMn}$ = 6.0 and 8.0 nm, respectively. As for the non-biased case (Fig. 4a), in which no loop is observed at $H_X$ = 0, application of positive (negative) $H_X$ results in clockwise (counter clockwise) loops. This result is again consistent with the finding that PtMn has a positive spin Hall angle. The same trend is seen for the biased sample (Fig. 4b); however, the boundary between the clockwise and counter-clockwise loops moves to $\mu_0 H_X$ = −10 mT due to the exchange bias. Critical current $I_C$ is defined as $I_{CH}$ at which $R_H$ changes by more than half (average for up-to-down and down-to-up switching), and plotted as a function of $H_X$ in Fig. 4c,d. As predicted by a theory[6], $I_C$ is proportional to −$H_X$. $H_X$ and $I_C$ at which the linear fits crosses are defined as $H_X^0$ and $I_C^0$, respectively, which are obtained to be −0.4±2.0 mT and 13.2±0.4 mA for $t_{PtMn}$ = 6.0 nm and −8.3±1.4 mT and 24.2±0.4 mA for $t_{PtMn}$ = 8.0 nm.

Figure 5a,b shows the $t_{PtMn}$ dependence of $H_X^0$ and $I_C^0$, respectively. $H_{bias}$ determined from the measured $m$-$H_X$ loops is also plotted in Fig. 5a, and the critical current density $J_C^0$ corresponding to $I_C^0$, which is calculated using a separately evaluated current flow ratio (see Supplementary Information), is plotted in Fig. 5b. $H_X^0$ varies in accordance with $H_{bias}$ (Fig. 5a), quantitatively evidencing that the observed field-free switching is attributed to the exchange-bias. $J_C^0$ is in the order of $10^{10}$ A m$^{-2}$, which is again comparable to the reported values for Pt/Co/AlO$_X$[3] and Ta/CoFeB/MgO[2,10] systems with a similar device geometry, indicating that PtMn possesses a spin Hall angle of the order of +0.1. Also, one can see that $I_C^0$ and $J_C^0$ discontinuously changes at $t_{PtMn}$ = 7.0 nm, which coincides with the thickness above which the exchange bias arises. According to a macrospin-based theory[6], $J_C^0$ depends on the saturation magnetization, thickness, and effective anisotropy field of the ferromagnetic layer and effective spin Hall angle of the channel. In the present system, the effective spin Hall angle should be almost independent of $t_{PtMn}$, since the spin diffusion length was found to be very short (= 0.5±0.1 nm)[24]. Also, we find that the saturation magnetization and effective anisotropy field are virtually independent of $t_{PtMn}$ (see Supplementary Information). Therefore, the observed $t_{PtMn}$ dependence of $J_C^0$ cannot be explained by the macrospin picture. The reason will be discussed below.

**Discussion on memristive nature.** As shown above, the perpendicular component of magnetization in the present PtMn/[Co/Ni] system, which can be controlled solely by an electric current, changes in an analogue manner according to current magnitude. Here we discuss the origin and technological significances of this phenomenon.

The intermediate $R_H$ value should be attributed to a multi-domain state. We also find that it cannot be observed for the case of the non-exchange-biased devices (see Supplementary Information), suggesting that the multi-domain state is related to the exchange bias. The multi-domain state is generally formed in case that a domain wall propagation field is larger than a nucleation field. (If the former is smaller than the latter, the domain wall immediately propagates after the nucleation takes place, as observed in the case of a Ta/CoFeB/MgO system[18]). In fact, it is shown by the dependence of the *m-H* loops on magnetic-field angle that the domain wall propagation in continuous films with exchange bias hardly occurs and that nucleation-mediated reversal is more dominant (see Supplementary Information). Thus, the exchange bias in the present system should play a key role in the observed analogue-like behaviour. We also perform a magnetic domain observation using magneto-optic Kerr effect microscope with 1-μm-spatial resolution after forming the intermediate state; however, no domain pattern is observed, indicating that the domain period is smaller than 1 μm. This fact suggests that the length scale of the reversed domain is not determined by the Kaplan–Gehring model which generally holds true for thin magnetic films[34], because, following the model, the domain period in the present system is derived to be in the order of 1 – 10 μm.

For thorough understanding of the mechanism for forming the multi-domain state, further investigation should be required. However, one possible explanation is that the multi-domain structure is formed with a length scale of the crystalline grains (~ 10 nm), among which the domain wall hardly propagates. Since our samples have a polycrystalline structure with a fiber texture (where the easy axes of AFM grains lie in the film plane with a two-dimensional random distribution and aggregation of the grains exhibits the exchange bias in a collective manner[27]), the SOT-induced switching is likely to occur in the grains in which the easy axis direction of the AFM is close to the channel direction. Once the magnetization is reversed in such grains, the domain wall cannot propagate to the neighbouring domains because the energy cost of forming a domain wall across the domains becomes high due to the different in-plane easy axes between the domains. As a result, the magnetization reversal proceeds in units of crystalline grains as the applied current increases, and stable multi-domain state is thus formed during the process. This also explains the fact that $J_C^0$ become smaller for non-exchange-biased devices (Fig. 5b), because the formed domain walls are free in the film plane and can easily propagate in this case.

A circuit element that can *memorize* the amount of past electric charge as its *resistance* is called as *memristor*[28], which is known to be useful for realizing neuromorphic computing[32], because it can perform the function of synapses. Much effort has been made to realize a memristor based on solid-state devices[29-31]. Since the magnetization state, which manifests in the resistance value in a real three-terminal device, can be continuously modulated by the magnitude of applied current as shown in Fig. 3b-e, our AFM/FM devices also serve just as a memristor.

In conclusion, this study sheds light on a new possibility of antiferromagnetic materials. It is found that antiferromagnetic PtMn shows the SHE with magnitude comparable to that of nonmagnetic Ta and Pt, in addition to the exchange bias. The two effects lead to a SOT-induced switching of perpendicular magnetization at zero fields in PtMn/[Co/Ni] bilayer system. Furthermore, the memristive nature arises as the exchange bias increases. The AFM/FM system thus offers a new avenue for electrical manipulation of magnetization and its application for solid-state devices and neuromorphic computing.

**Methods**

**Film preparation.** The films were deposited at room temperature onto 3-inch high-resistivity Si wafers with a natural oxidation layer. RF magnetron sputtering was used to deposit the MgO layer, and DC magnetron sputtering was used to deposit the other layers. Base pressure of the chamber was less than $1\times10^{-6}$ Pa, and Ar gas was used for sputtering. No magnetic field was applied during the sputtering. Composition of PtMn is $Pt_{40}Mn_{60}$ (atomic %). After the deposition, the samples were annealed at 300 °C for two hours under an in-plane magnetic field of 1.2 T in order to provide the exchange bias.

**Characterization of blanket film.** The *m-H* loops were measured using a vibrating sample magnetometer, where *m* denotes magnetic moment per unit area. Bias field $H_{bias}$ was determined from the shift of the *m-$H_X$* loop from the origin; in more detail, the average of $H_X$ at which the sign of *m* changes was taken. The effective anisotropy field was determined from the difference between the *m-H* loops in the *Z* and *Y* directions. Sheet resistance was measured by using a standard dc four-probe method.

**Device fabrication.** The deposited films were processed into Hall devices. First a stack (Ta/ Pt/ PtMn/ [Co/Ni]$_2$/ Co/ MgO/ Ta) was patterned into a cross shape by photolithography and Ar ion milling. Then, electrodes and contact pads made of Cr(5 nm)/ Au(100 nm) were formed by photolithography and lift-off. Width and length of the channel of the Hall devices were 10 and 100 μm, respectively, and those of the Hall probe were 3 and 30 μm, respectively.

**Electrical measurement.** All electrical measurements were performed at room temperature. $R_H$ was measured by supplying $I_{CH}$ from a dc current source while monitoring the Hall voltage by a nano-voltmeter. See Fig. 1d for the definition of the sign of each terminal. $R_H$-$H_Z$ loops shown in Figs. 1e and 2ab were obtained by measuring $R_H$ while sweeping $H_Z$. The $R_H$-$I_{CH}$ loops at zero fields shown in Fig. 3b-d were obtained with the sequence shown in Fig. 3a. First, a current pulse for initialization with the magnitude of $I_{INIT}$ and duration of 0.5 s was applied. Then, pulsed $I_{CH}$ application followed by $R_H$ measurement was repeated while the magnitude of $I_{CH}$ was varied. The duration of the pulsed $I_{CH}$ was 0.5 s. $R_H$ was measured with $I_{CH}$ of 1 mA. The $R_H$-$I_{CH}$ loops under $H_X$ shown in Fig. 4ab were obtained as follows. Static $H_X$ was applied first and then pulsed $I_{CH}$ application followed by $R_H$ measurement was repeated while $I_{CH}$ was varied from −35 to +34 mA and then from +35 to −34 mA.

**References**

1  Miron, I. M. *et al.* Perpendicular switching of a single ferromagnetic layer induced by in-plane current injection. *Nature* **476**, 189-193 (2011).

2  Liu, L. *et al.* Spin-torque switching with the giant spin Hall effect of tantalum. *Science* **336**, 555-558 (2012).

3  Liu, L., Lee, O. J., Gudmundsen, T. J., Ralph, D. C. & Buhrman, R. A. Current-Induced Switching of Perpendicularly Magnetized Magnetic Layers Using Spin Torque from the Spin Hall Effect. *Phys. Rev. Lett.* **109** (2012).

4  Pai, C.-F. *et al.* Spin transfer torque devices utilizing the giant spin Hall effect of tungsten. *Appl. Phys. Lett.* **101**, 122404 (2012).

5  Kim, J. *et al.* Layer thickness dependence of the current-induced effective field vector in Ta|CoFeB|MgO. *Nature Mater.* **12**, 240-245 (2013).

6   Lee, K.-S., Lee, S.-W., Min, B.-C. & Lee, K.-J. Threshold current for switching of a perpendicular magnetic layer induced by spin Hall effect. *Appl. Phys. Lett.* **102**, 112410 (2013).

7   Yamanouchi, M. *et al.* Three terminal magnetic tunnel junction utilizing the spin Hall effect of iridium-doped copper. *Appl. Phys. Lett.* **102**, 212408 (2013).

8   Garello, K. *et al.* Symmetry and magnitude of spin-orbit torques in ferromagnetic heterostructures. *Nature Nanotech.* **8**, 587-593 (2013).

9   Zhang, C. *et al.* Magnetotransport measurements of current induced effective fields in Ta/CoFeB/MgO. *Appl. Phys. Lett.* **103**, 262407 (2013).

10  Zhang, C. *et al.* Magnetization reversal induced by in-plane current in Ta/CoFeB/MgO structures with perpendicular magnetic easy axis. *J. Appl. Phys.* **115**, 17C714 (2014).

11  Lee, K.-S., Lee, S.-W., Min, B.-C. & Lee, K.-J. Thermally activated switching of perpendicular magnet by spin-orbit spin torque. *Appl. Phys. Lett.* **104**, 072413 (2014).

12  Pai, C.-F. *et al.* Enhancement of perpendicular magnetic anisotropy and transmission of spin-Hall-effect-induced spin currents by a Hf spacer layer in W/Hf/CoFeB/MgO layer structures. *Appl. Phys. Lett.* **104**, 082407 (2014).

13  Fan, Y. *et al.* Magnetization switching through giant spin-orbit torque in a magnetically doped topological insulator heterostructure. *Nature Mater.* **13**, 699-704 (2014).

14  Yu, G. *et al.* Switching of perpendicular magnetization by spin-orbit torques in the absence of external magnetic fields. *Nature Nanotech.* **9**, 548-554 (2014).

15  Yu, G. *et al.* Current-driven perpendicular magnetization switching in Ta/CoFeB/[TaOx or MgO/TaOx] films with lateral structural asymmetry. *Appl. Phys. Lett.* **105**, 102411 (2014).

16  Garello, K. *et al.* Ultrafast magnetization switching by spin-orbit torques. *Appl. Phys. Lett.* **105**, 212402 (2014).

17  Akyol, M. *et al.* Current-induced spin-orbit torque switching of perpendicularly magnetized Hf|CoFeB|MgO and Hf|CoFeB|TaOx structures. *Appl. Phys. Lett.* **106**, 162409 (2015).

18 Zhang, C., Fukami, S., Sato, H., Matsukura, F. & Ohno, H. *Appl. Phys. Lett.*, in press (2015).

19 Fukami, S., Yamanouchi, M., Ikeda, S. & Ohno, H. Domain wall motion device for nonvolatile memory and logic - Size dependence of device properties. *IEEE Trans. Magn.* **50**, 3401006 (2014).

20 Hoffmann, A. Spin Hall effect in metals. *IEEE Trans. Magn.* **49**, 5172 (2013).

21 Sinova, J., Valenzuela, S. O., J., W., Back, C. H. & Jungwirth, T. Spin Hall effect. Preprint at http://arxiv.org/abs/1411.3249 (2014).

22 Chen, H., Niu, Q. & MacDonald, A. H. Anomalous Hall Effect Arising from Noncollinear Antiferromagnetism. *Phys. Rev. Lett.* **112** 017205 (2014).

23 Mendes, J. B. S. *et al.* Large inverse spin Hall effect in the antiferromagnetic metal $Ir_{20}Mn_{80}$. *Phys. Rev. B* **89**, 140406(R) (2014).

24 Zhang, W. *et al.* Spin Hall Effects in Metallic Antiferromagnets. *Phys. Rev. Lett.* **113**, 196602 (2014).

25 Meiklejohn, W. H. & Bean, C. P. New Magnetic Anisotropy. *Phys. Rev.* **102**, 1413 (1956).

26 Stiles, M. D. & McMichael, R. D. Model for exchange bias in polycrystalline ferromagnet-antiferromagnet bilayers. *Phys. Rev. B* **59**, 3722 (1999).

27 Tsunoda, M. & Takahashi, M. Field independent rotational hysteresis loss on exchange coupled polycrystalline Ni–Fe/Mn–Ir bilayers. *J. Appl. Phys.* **87**, 6415 (2000).

28 Chua, L. O. Memristor -The missing circuit element. *IEEE Trans. Circuit Theory* **18**, 507 (1971).

29 Strukov, D. B., Snider, G. S., Stewart, D. R. & Williams, R. S. The missing memristor found. *Nature* **453**, 80-83 (2008).

30 Jo, S. H. *et al.* Nanoscale memristor device as synapse in neuromorphic systems. *Nano Lett.* **10**, 1297-1301 (2010).

31 Locatelli, N., Cros, V. & Grollier, J. Spin-torque building blocks. *Nature Mater.* **13**, 11-20 (2014).

32  Merolla, P. A. *et al.* A million spiking-neuron integrated circuit with a scalable communication network and interface. *Science* **345**, 668-673 (2014).

33  Suzuki, T. *et al.* Current-induced effective field in perpendicularly magnetized Ta/CoFeB/MgO wire. *Appl. Phys. Lett.* **98**, 142505 (2011).

34  Kaplan, B. & Gehring, G. A. The domain structure in ultrathin magnetic films. *J. Magn. Magn. Mater.* **128**, 111 (1993).

**Acknowledgements**

The authors thank C. Igarashi, T. Hirata, Y. Kawato, H. Iwanuma, and K. Goto for their technical support. A portion of this work was supported by the R&D Project for ICT Key Technology to Realize Future Society of MEXT, R&D Subsidiary Program for Promotion of Academia-industry Cooperation of METI, ImPACT Program of CSTI, and JSPS KAKENHI 15J04691.

**Author contributions**

S.F. and H.O. planned the study. S.F. deposited the film. S.F. and C.Z. fabricated the samples. S.F. performed the measurements and analyzed the data. S.D. and S.F. performed the MOKE microscopy observation. S.F. wrote the manuscript with input from H.O., C.Z., and S.D. All authors discussed the results.

**Figure Captions**

**Figure 1 | Sample structure and magnetic properties. a**, Schematic of the studied AFM/FM bilayer system with the definition of *X-Y-Z* coordinate. A ferromagnetic Co/Ni multilayer with a

perpendicular easy axis is located on top of an antiferromagnetic PtMn layer. Magnetization in the vicinity of the PtMn interface in the Co/Ni multilayer tilts in the +X direction due to the exchange bias. **b,c**, *m-H* loops with $t_{PtMn}$ = 6.0 and 8.0 nm, respectively. A magnetic field is applied in the *X*, *Y*, and *Z* directions. Insets show the loops for the magnetic field range of ±700 mT. Main panels show the loops in a narrow field range. **d**, Optical micrograph of the fabricated Hall device. **e,f**, $R_H$-$H_Z$ loops and corresponding *m-$H_Z$* loops, respectively, for $t_{PtMn}$ = 6.0, 7.5, 7.5, 8.0, and 8.5 nm.

**Figure 2 | $R_H$-$H_Z$ loops measured with various $I_{CH}$ for non-exchange-biased and exchange-biased structures. a,b**, $R_H$-$H_Z$ loops for $t_{PtMn}$ = 6.0 and 8.0 nm, respectively. Those obtained with $I_{CH}$ = −32, +1, +32 mA are shown. **c,d**, Coercive field $H_C$ as a function of $I_{CH}$ for $t_{PtMn}$ = 6.0 and 8.0 nm, respectively. $H_C$ for both down-to-up and up-to-down reversals is retracted from the $R_H$-$H_Z$ loops shown in **a** and **b**. **e,f**, Schematics showing effective fields of Slonczewski-like SOT (**$H_{SL}$**) acting on magnetization (**M**) with the flow of spin-polarized electrons under the presence of the SHE in PtMn for $t_{PtMn}$ = 6.0 and 8.0 nm, respectively. Clockwise **$H_{SL}$** (blue arrows) around the *Y* axis explains well the results shown in **c,d**.

**Figure 3 | $R_H$-$I_{CH}$ loops at zero fields. a**, Procedure for measuring $R_H$-$I_{CH}$ loops. First, the initialization pulse with the magnitude of $I_{INIT}$ is applied to the channel. Then, pulsed $I_{CH}$ with duration of 0.5 s and various magnitudes is applied. After every pulse application, $R_H$ is measured (indicated by red triangles). **b-e**, Obtained $R_H$-$I_{CH}$ loops for $t_{PtMn}$ = 8.0 nm. **b,c**, $R_H$-$I_{CH}$ loops for devices with exchange bias in the +*X* direction, where clockwise loops are obtained. **d,e**, $R_H$-$I_{CH}$ loops for devices with exchange bias in the −*X* direction, where counter clockwise loops are obtained. **b,e**, $R_H$-$I_{CH}$ loops obtained after a positive initialization pulse is applied.

Pulse applications of negative $I_{CH}$ up to $-I_{MAX}$ precede positive $I_{CH}$ up to $+I_{INIT}$. **c,d**, $R_H$-$I_{CH}$ loops obtained after negative initialization pulse is applied. Pulse applications of positive $I_{CH}$ up to $+I_{MAX}$ precede negative $I_{CH}$ up to $-I_{INIT}$. $I_{MAX}$ is varied from 8 to 44 mA. Arrows in panels indicate the position from which the measurement starts

**Figure 4 | $R_H$-$I_{CH}$ loops under various $H_X$. a,b**, $R_H$-$I_{CH}$ loops obtained under various $H_X$ for $t_{PtMn}$ = 6.0 and 8.0 nm, respectively. The magnitude of $\mu_0 H_X$ ranges from $-40$ to $+40$ mT. **c,d**, $I_C$ evaluated from the obtained $R_H$-$I_{CH}$ loops as a function of $H_X$. $I_C$ obtained from clockwise (CW) and counter clockwise (CCW) loops are plotted as different symbols. Solid lines are linear fits to the data.

**Figure 5 | $t_{PtMn}$ dependence of bias field and critical current (density). a**, $t_{PtMn}$ dependences of $H_X^0$ obtained from $R_H$-$I_{CH}$ loops and $H_{bias}$ obtained from $m$-$H_X$ loops. **b**, $t_{PtMn}$ dependences of $I_X^0$ obtained from $R_H$-$I_{CH}$ loops and corresponding $J_C^0$. Error bars come from the standard error of the linear fit shown in Fig. 4cd as solid lines.

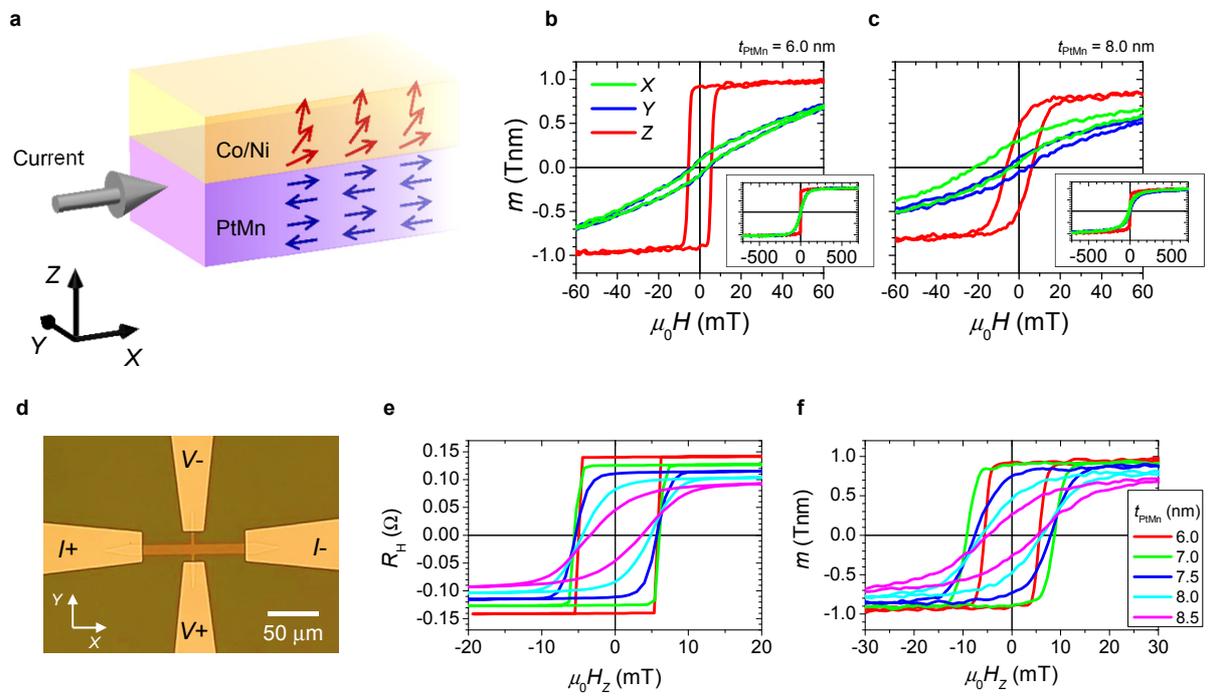

Figure 1, Fukami *et al*.

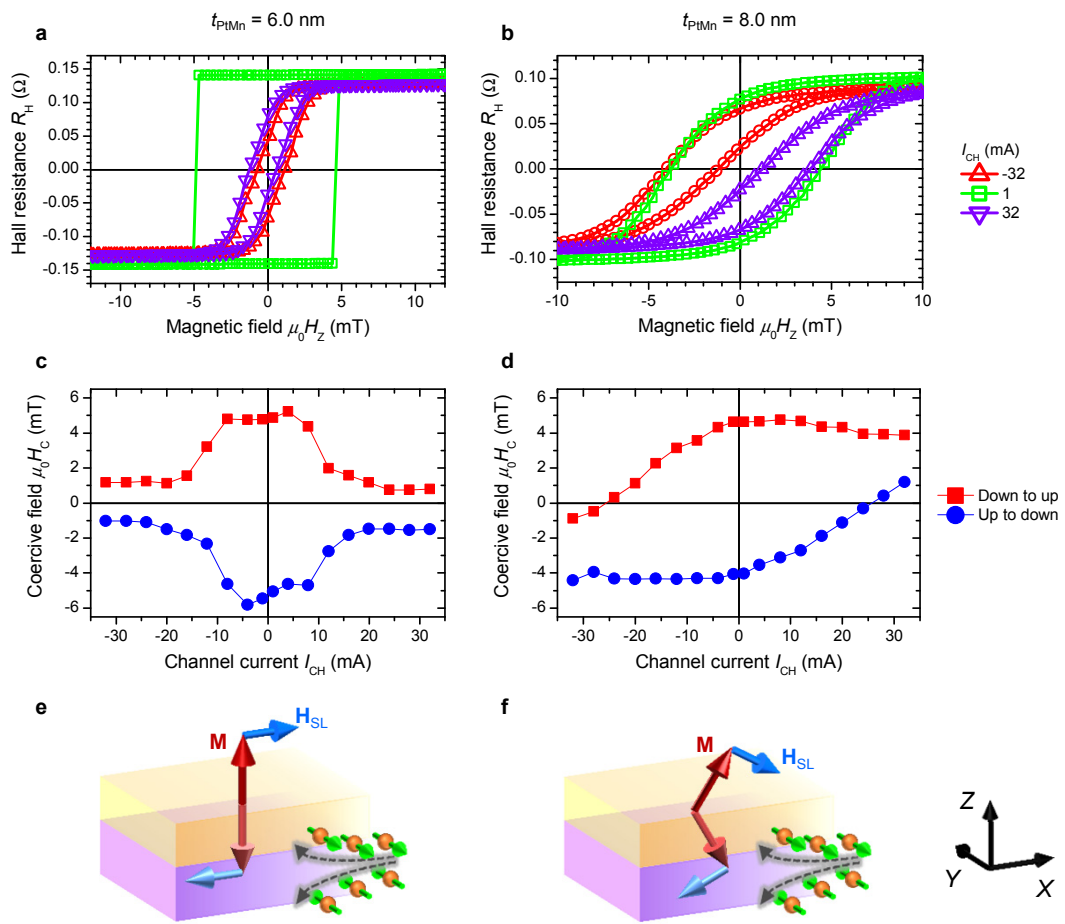

Figure 2, Fukami *et al*.

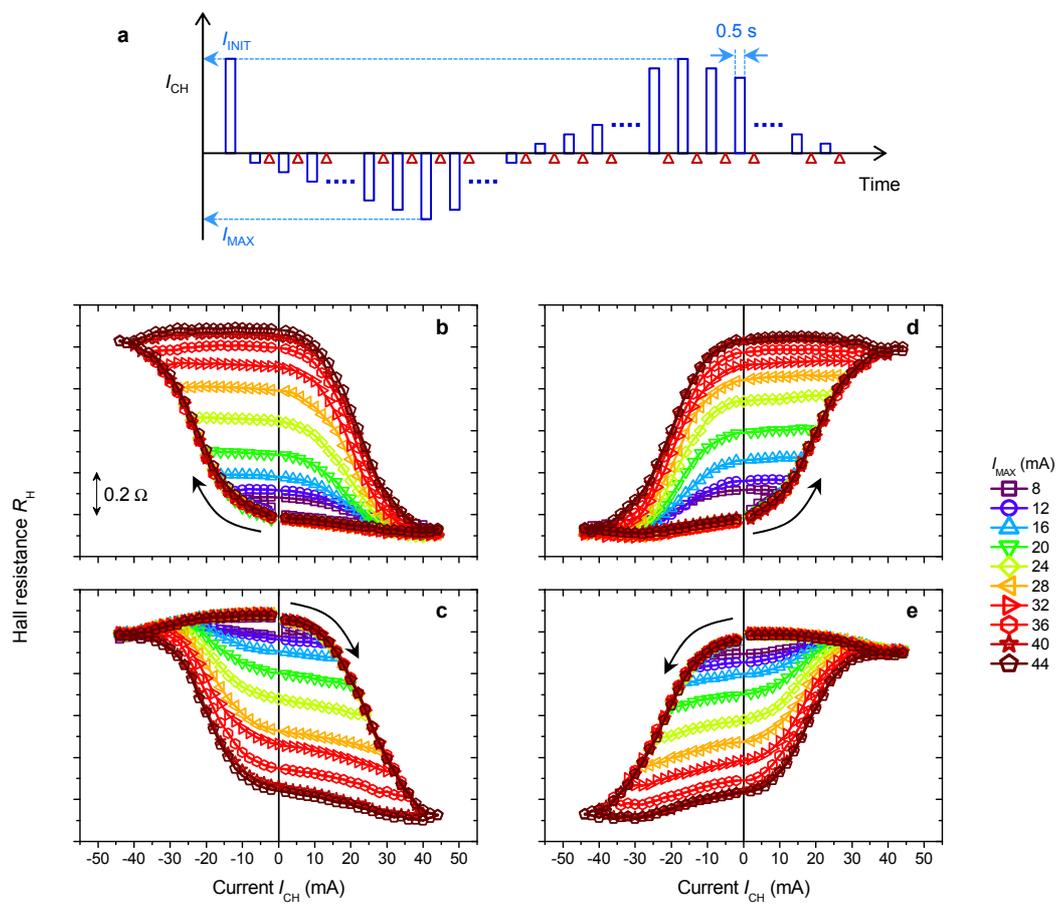

Figure 3, Fukami *et al*.

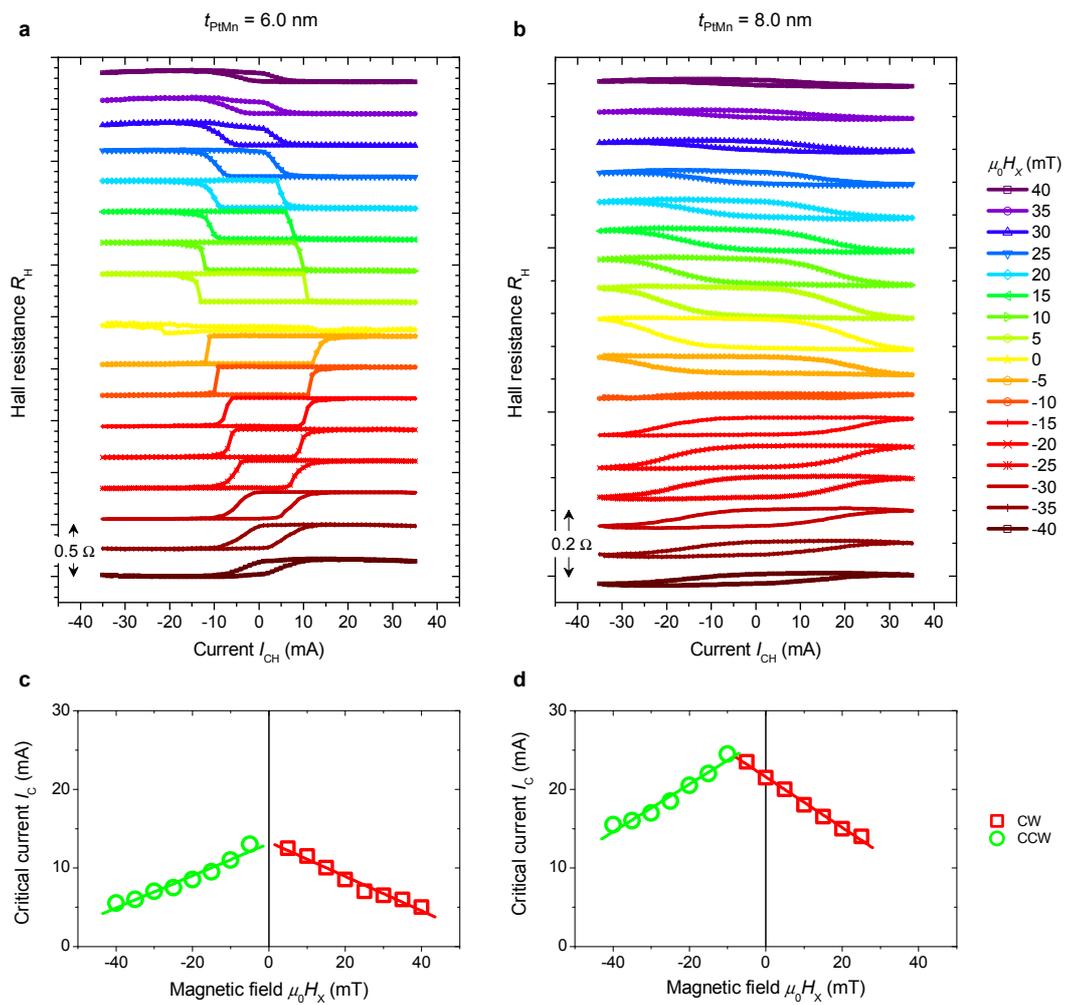

Figure 4, Fukami *et al*.

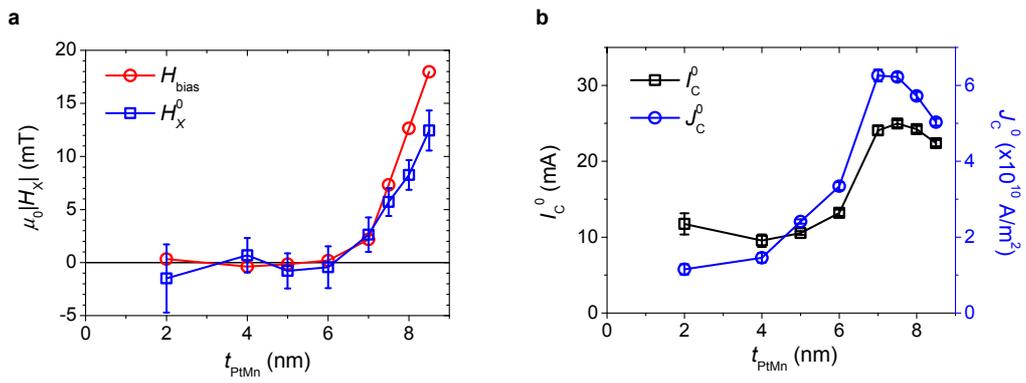

Figure 5, Fukami *et al*.

Supplementary Information

# Magnetization switching by spin-orbit torque in antiferromagnet/ferromagnet bilayer systems

S. Fukami[1,2], C. Zhang[3], S. DuttaGupta[3], and H. Ohno[1,2,3,4]

[1]*Center for Spintronics Integrated Systems, Tohoku University, 2-1-1 Katahira, Aoba, Sendai 980-8577, Japan*

[2]*Center for Innovative Integrated Electronic Systems, Tohoku University, 468-1 Aoba, Aramaki, Aoba, Sendai 980-0845 Japan*

[3]*Laboratory for Nanoelectronics and Spintronics, Research Institute of Electrical Communication, Tohoku University, 2-1-1 Katahira, Aoba, Sendai 980-8577, Japan*

[4]*WPI Advanced Institute for Materials Research, Tohoku University, 2-1-1 Katahira, Aoba, Sendai 980 8577, Japan*

**S1. *m-H* loops of blanket films and magnetic properties versus $t_{PtMn}$**

In the present study, we investigate the spin-orbit torque (SOT) induced magnetization switching of PtMn/[Co/Ni] structures with various PtMn thicknesses $t_{PtMn}$. Here we show the magnetization hysteresis loop (*m-H* loop) of the blanket films and $t_{PtMn}$ dependence of exchange bias field $H_{bias}$ and effective anisotropy field $H_K^{eff}$.

Figure S1a–h shows the *m-H* loops of the blanket films with $t_{PtMn}$ = 2.0, 4.0, 5.0, 6.0, 7.0, 7.5, 8.0, 8.5 nm, respectively. The loops obtained by applying magnetic fields in the *X*, *Y*, and *Z*

directions are shown in the figure (Cartesian coordinate system is defined in the main body). The saturation value of $m$ is around 1.1 T nm for all the $t_{PtMn}$, indicating that the saturation magnetization does not depend on $t_{PtMn}$. $m$-$H_X$ loop (green curves) is overlapped with $m$-$H_Y$ loop (blue curves) for $t_{PtMn} \leq 6$ nm, indicating no exchange bias, whereas the former is shifted to $-H_X$ direction for $t_{PtMn} \geq 7$ nm due to the exchange bias in the $+X$ direction. Also, a decrease in the squareness of $m$-$H_Z$ loop (red curves) accompanies with the increase in the exchange bias. We note that the perpendicular easy axis cannot be obtained by further increase in $t_{PtMn}$.

Figure S1i shows the $H_{bias}$ and $H_K^{eff}$ determined from the $m$-$H$ loops shown in Fig. S1a–h. See Method in the main body for how to determine the $H_{bias}$ and $H_K^{eff}$. $H_{bias}$ is effectively zero for $t_{PtMn} \leq 6$ nm, whereas it increases with $t_{PtMn}$ for $t_{PtMn} \geq 7$ nm. $H_K^{eff}$ shows almost constant value in the range of 120 – 150 mT.

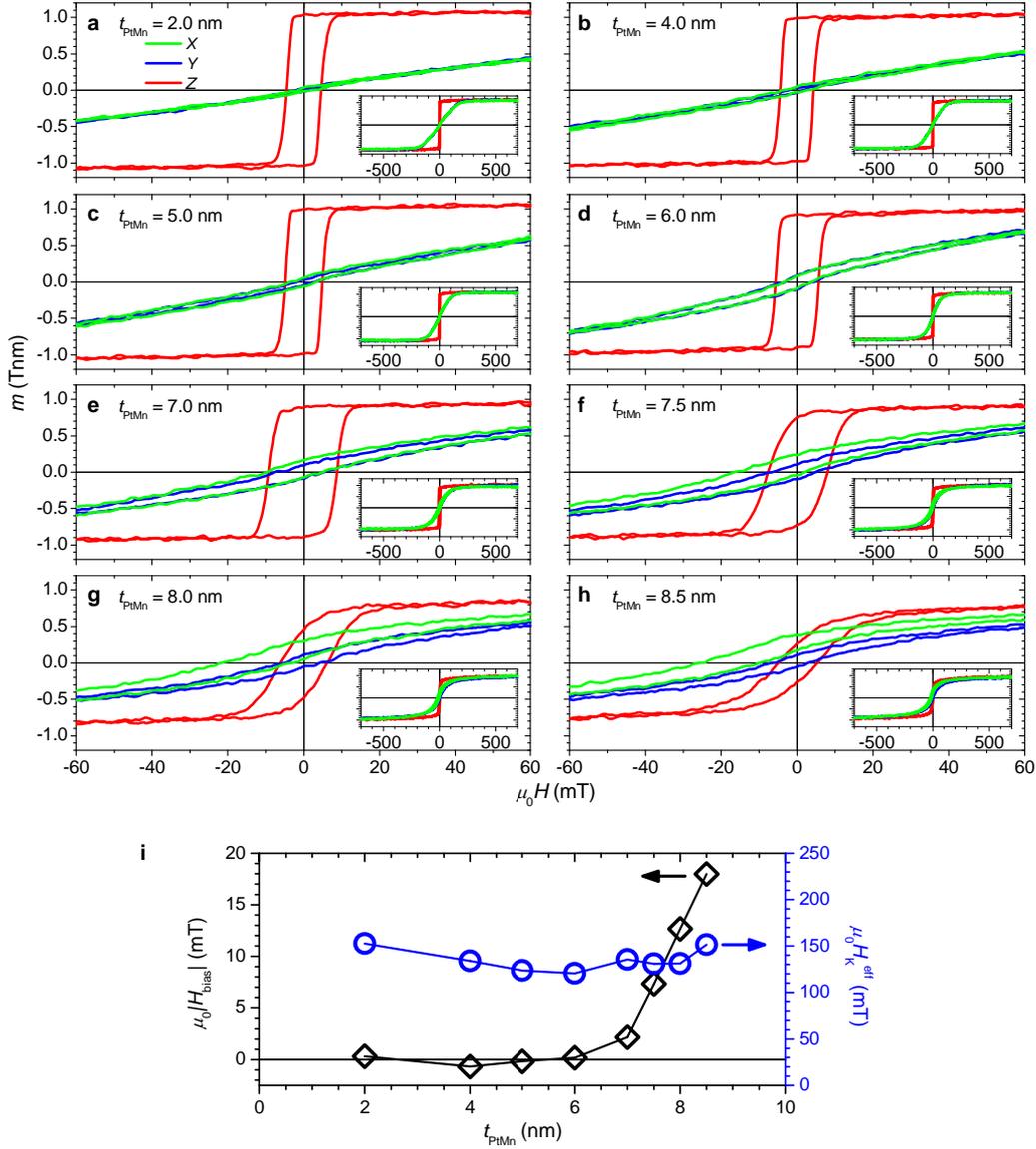

**Figure S1 | Magnetic properties of blanket films. a-h**, $m$-$H$ loops of the blanket films with $t_{PtMn}$ = 2.0, 4.0, 5.0, 6.0, 7.0, 7.5, 8.0, 8.5 nm, respectively. Insets show the loops for the magnetic field range of ±700 mT, whereas the main panels show the loops in a low field range. **i**, $t_{PtMn}$ dependence of $H_{bias}$ and $H_K^{eff}$.

## S2. Sheet resistance and current flow ratio

We here show the transport properties of blanket films. Figure S2a shows the sheet resistance of PtMn/[Co/Ni] stacks studied here as a function of $t_{PtMn}$. The sheet resistance is measured using a standard four-probe method. It decreases from 85.8 to 66.7 Ω with increasing $t_{PtMn}$ from 0 to 8.5 nm. In order to determine the current flow ratio in each layer, we measure the sheet resistance of various stacks with the following structures:

1. Sub./ Ta(3)/ Pt(4)/ PtMn(8.0)/ [Co(0.3)/ Ni(0.6)]$_2$/ Co(0.3)/ MgO(1.2)/ Ta(2)
2. Sub./ Ta(3)/ Pt(4)/ PtMn(8.0)/ [Co(0.3)/ Ni(0.6)]$_2$/ Co(0.3)/ MgO(1.2)/ Ta(0.5)
3. Sub./ Ta(3)/ Pt(4)/ PtMn(8.5)/ MgO(1.2)/ Ta(2)
4. Sub./ Ta(3)/ Pt(4)/ PtMn(8.0)/ MgO(1.2)/ Ta(2)
5. Sub./ Ta(3)/ Pt(4)/ PtMn(7.5)/ MgO(1.2)/ Ta(2)
6. Sub./ Ta(3)/ Pt(4)/ PtMn(7.0)/ MgO(1.2)/ Ta(2)
7. Sub./ Ta(3)/ Pt(4)/ PtMn(6.0)/ MgO(1.2)/ Ta(2)
8. Sub./ Ta(3)/ Pt(4)/ PtMn(5.0)/ MgO(1.2)/ Ta(2)
9. Sub./ Ta(3)/ Pt(4)/ PtMn(4.0)/ MgO(1.2)/ Ta(2)
10. Sub./ Ta(3)/ Pt(4)/ PtMn(2.0)/ MgO(1.2)/ Ta(2)
11. Sub./ Ta(3)/ Pt(4)/ MgO(1.2)/ Ta(2)
12. Sub./ Ta(3)/ MgO(1.2)/ Ta(2)

Assuming that the total resistance can be expressed by the summation of parallel resistance of each layer, we derive the sheet resistance of the individual layers from the measured sheet resistance of above 12 samples. Then, the current flow ratio in each layer is obtained as shown in Figure S2b. As $t_{PtMn}$ increases from 0 to 8.5 nm, the current flow ratio in the PtMn layer increases from 0 to 19.6%. The obtained current flow ratio is used when converting $I_{CH}$ to the corresponding current density in PtMn layer.

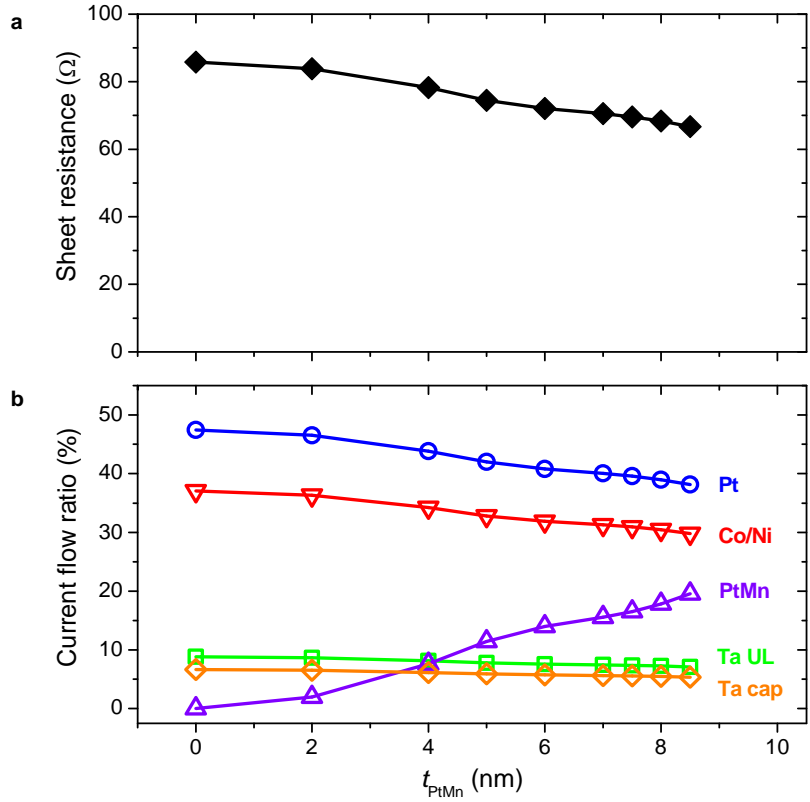

**Figure S2 | Sheet resistance and current flow ratio. a**, Sheet resistance of the stacks with Si sub./ Ta(3)/ Pt(4)/ PtMn($t_{PtMn}$)/ [Co(0.3)/Ni(0.6)]$_2$/ Co(0.3)/ MgO(1.2)/ Ta(2) as a function of $t_{PtMn}$. **b**, Calculated current flow ratio in each layer as a function of $t_{PtMn}$.

## S3. $R_H$-$H_Z$ loops with various $I_{CH}$ for opposite exchange bias

In the main body, we show in Fig. 2b the $R_H$-$H_Z$ loops of a device with $t_{PtMn}$ = 8.0 nm, which was annealed at 300°C under a magnetic field of +1.2 T in the $X$ direction. Here we show the $R_H$-$H_Z$ loop of a device with the same stack structure and opposite exchange bias direction, which is prepared by annealing under the magnetic field of –1.2 T in the $X$ direction. Figure S3a shows the obtained $R_H$-$H_Z$ loops for $I_{CH}$ = –32, +1, +32 mA. The $R_H$-$H_Z$ loop for $I_{CH}$ = 1 mA is virtually the same with the one shown in Fig. 2b, indicating that the magnitude of magnetic anisotropy and exchange bias are equivalent between the two devices with opposite bias direction. On the other hand, the loop for $I_{CH}$ = –32 (+32) mA in Fig. S3a is similar to the one for $I_{CH}$ = +32 (−32) mA in Fig. 2b. Figure S3b shows the coercive field $H_C$ as a function of $I_{CH}$. One can see a mirror symmetry tendency with respect to the sign of $I_{CH}$ between the Fig. 2d and Fig. S3b. This is consistent with the scenario that the Slonczewski-like SOT and exchange bias govern the magnetization reversal. In the present case, since the magnetization of Co/Ni layer tilts to −X direction, a positive current that generates clockwise $\mathbf{H}_{SL}$ around Y axis (see Fig. 2f) promotes down-to-up reversal, whereas a negative current that generates counter clockwise $\mathbf{H}_L$ promotes up-to-down reversal, which are in agreement with what we see in Fig. S3b.

From the results shown in Fig. 2d and Fig. S3b, we derive a current-induced effective field. Since the $H_C$ appears to be proportional to the $I_{CH}$, which assists the reversal, we fit a linear function to the relation between $H_C$ and $I_{CH}$. In more detail, we perform the fitting for the following four regions: (i) up-to-down reversal for positive $I_{CH}$ in Fig. 2d, (ii) down-to-up reversal for negative $I_{CH}$ in Fig. 2d, (iii) up-to-down reversal for negative $I_{CH}$ in Fig. S3b, and (iv) down-to-up reversal for positive $I_{CH}$ in Fig. S3b. The slopes give the effective field per unit current, which are obtained to be 0.169, 0.191, −0.176, −0.196 T/A, respectively. The average and standard deviation of the absolute values are 0.183 and 0.012 T/A, respectively, which correspond to 77.6 and 5.2 mT per the current density of $10^{12}$ A m$^{-2}$. Note here that the obtained effective field here includes the effect of field-like SOT in addition to the Slonczewski-like SOT.

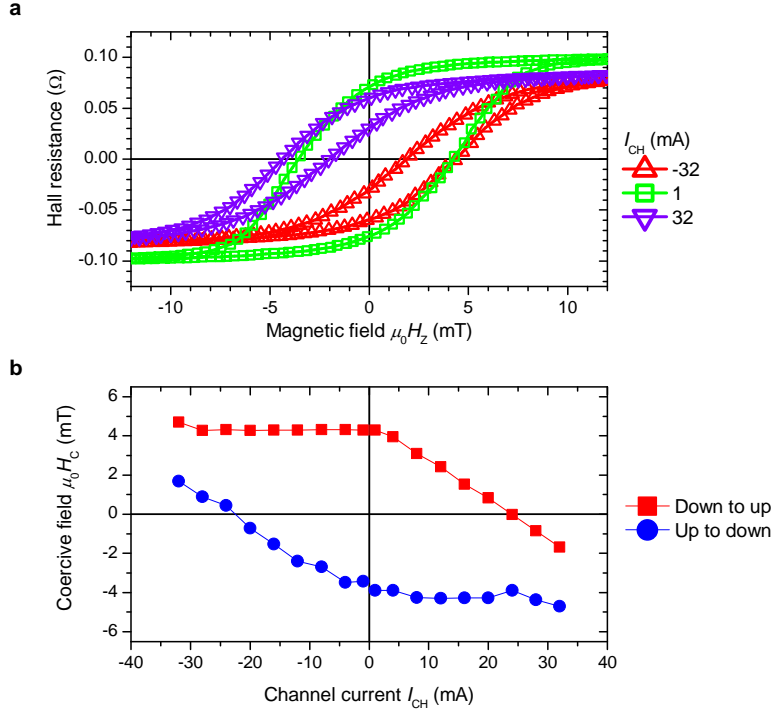

**Figure S3 | $R_H$-$H_Z$ loops for opposite exchange bias. a**, $R_H$-$H_Z$ loops for $t_{PtMn}$ = 8.0 nm measured with $I_{CH}$ = −32, +1, +32 mA. The exchange bias direction is opposite to the one used for Fig. 2b. **b**, Coercive field $H_C$ for down-to-up and up-to-down reversals as a function of $I_{CH}$.

## S4. $R_H$-$I_{CH}$ loops at no field for various $t_{PtMn}$

In the main body, we show the $R_H$-$I_{CH}$ loops measured at no field for a device with $t_{PtMn}$ = 8.0 nm (Fig. 3). Here we show the results obtained from other devices with different $t_{PtMn}$. Figure S4a–d shows the $R_H$-$I_{CH}$ loops for the devices with $t_{PtMn}$ = 7.0, 7.5, 8.0, 8.5 nm, respectively, where the magnitude of exchange bias is found to increase in this order (see Supplementary Information S1). The measurement sequence is the same to the one used in the main body (see Fig. 3a). The device studied here is annealed under the magnetic field of +$X$ direction. Positive initialization pulse is first applied and then pulses with negative $I_{CH}$ are applied up to −$I_{MAX}$ followed by the application of positive $I_{CH}$ up to +44 mA.

As can be seen, with increasing $t_{PtMn}$, the $R_H$ is more likely to take intermediate values. For example, for $t_{PtMn}$ = 7.0 nm (Fig. S4a), the $R_H$ after the $I_{CH}$ of less than −24 mA is applied goes back to its original value, whereas for $t_{PtMn}$ = 8.0 and 8.5 nm (Fig. S4d), the $R_H$ takes different values according to $I_{MAX}$ in the range of −8 to −36 mA. This means that the device with smaller exchange bias is more suitable to be used as a digital memory, whereas the one with larger exchange bias is more suitable to be used as a memristor. This behavior may be explained by the difference in how easily the domain walls propagate. As discussed in the main body, stronger exchange bias interrupts the domain wall propagation across the grain boundaries, resulting in the stable multi-domain state and memristive behavior.

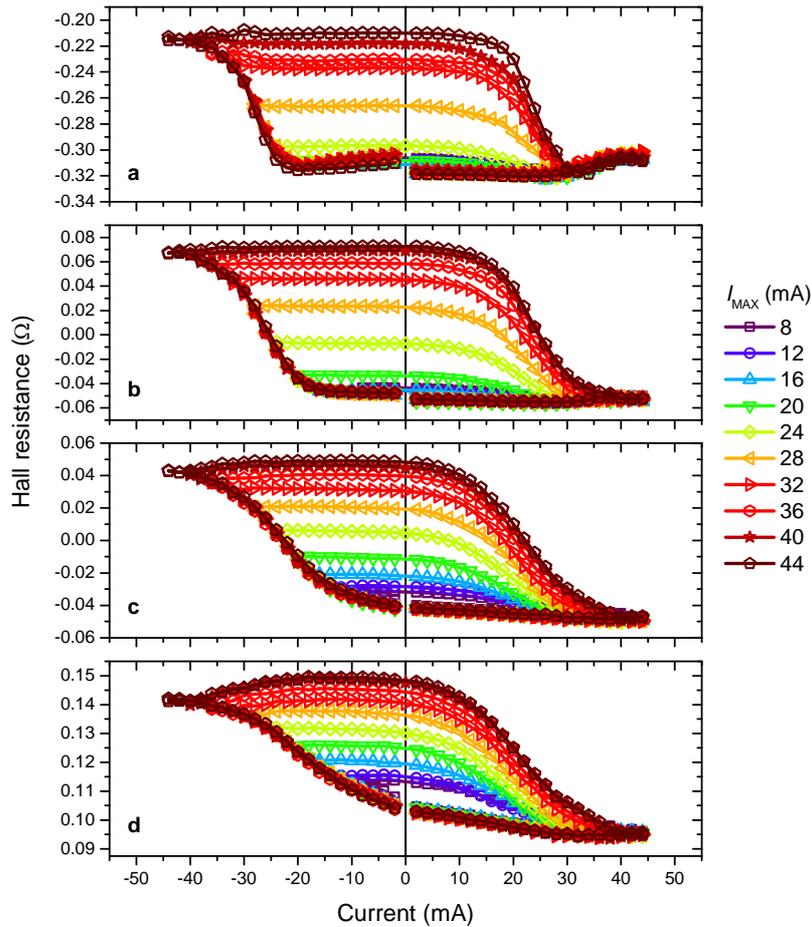

**Figure S4 | $R_H$-$I_{CH}$ loops for different strengths of exchange bias. a,b,c,d,** The loop for $t_{PtMn}$ = 7.0, 7.5, 8.0, 8.5 nm, respectively.

**S5. Demagnetizing loops of Hall device**

As shown in Figs. 3 and S4, the exchange-biased devices take intermediate states through the application of current. In order to clarify the origin of the intermediate state, we perform a demagnetization procedure using the Hall device with $t_{PtMn}$ = 8.0 nm (exchange-biased device). Figure S4a shows the $R_H$ as a function of perpendicular magnetic field $H_Z$. We serially sweep $H_Z$, while monitoring $R_H$, as follows; $+H_1$, $-H_2$, $+H_3$, $-H_4$, …, $+H_i$, $-H_{i+1}$, …, where $H_{i+1}$ is smaller than $H_i$. The demagnetization procedure gradually decreases the remanence down to almost zero, where a multi-domain state should be realized. Since the $R_H$ represents the magnetization state in the vicinity of the Hall cross whose size is $10 \times 3$ μm$^2$ in our samples, the obtained results indicate that the domain period is much smaller than this scale, which is smaller than the size expected from Kaplan–Gehring model[1]. We perform the same procedure to the device with $t_{PtMn}$ ≤ 6 nm (non-exchange-biased device), and also Ta/CoFeB/MgO samples we previously studied[2]; however the demagnetized state is not obtained by whether means. Thus, we conclude that the obtained intermediate $R_H$ value has something to do with the exchange bias.

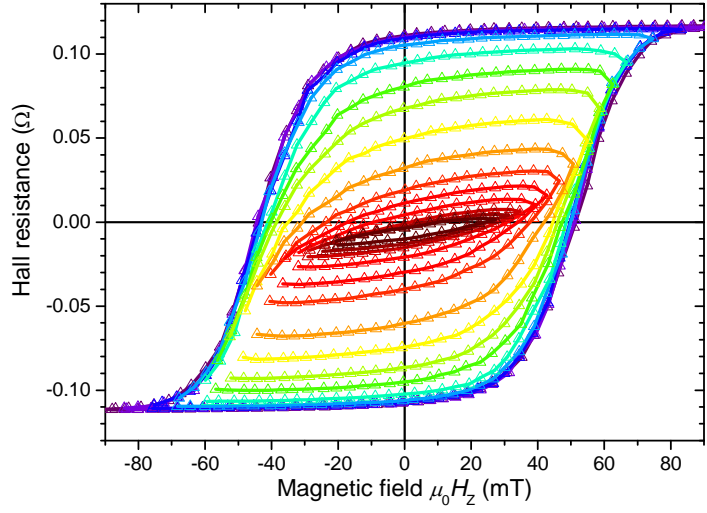

**Figure S5 | Demagnetization curves of an exchange-biased Hall device.** $R_H$ as a function of decaying magnetic field $H_z$ for the Hall device with $t_{PtMn}$ = 8.0 nm.

### S6. Magnetic field angle dependence of *m-H* loops

To investigate the magnetization reversal mode, we evaluate the magnetic field angle dependence of *m-H* loops of the blanket films. Here the blanket films with $t_{PtMn}$ = 0, 2, 4, 6, 8 nm and the size of 10 × 10 mm$^2$ are tested. The *m-H* loops are measured using a vibrating sample magnetometer with various magnetic field directions in the *Y-Z* plane. Figure S6a–e shows the *m-H* loops for $t_{PtMn}$ = 0, 2, 4, 6, 8 nm, respectively, measured by applying magnetic field along various directions. The direction of magnetic field is tilted from *Z* axis by $\theta$ = 0, 15, 30, 45, 60, and 75°. Figure S6f shows the switching field $H_{SW}$ as a function of $\theta$. It has been known that in case that the magnetization reversal is governed by a domain wall propagation, the switching field $H_{SW}$ is described by the Kondorsky model[3]: $H_{SW} \propto 1/\cos\theta$, which is shown as dotted line in the figure. The $\theta$ dependence of $H_{SW}$ is close to the Kondorsky model for $t_{PtMn}$ = 0, 2, 4, 6 nm, whereas that for $t_{PtMn}$ = 8 nm largely deviates from it. In general, the blanket films contain the reversed domain whose expansion through domain wall propagation determines the $H_{SW}$, and the Kondorsky model eventually holds true. Therefore, the deviation from the Kondorsky model seen in the films with $t_{PtMn}$ = 8 nm suggests that the domain wall propagation in the exchange-biased film is more unlikely to occur than the non-exchange-biased films, and nucleation-mediated reversal is more dominant.

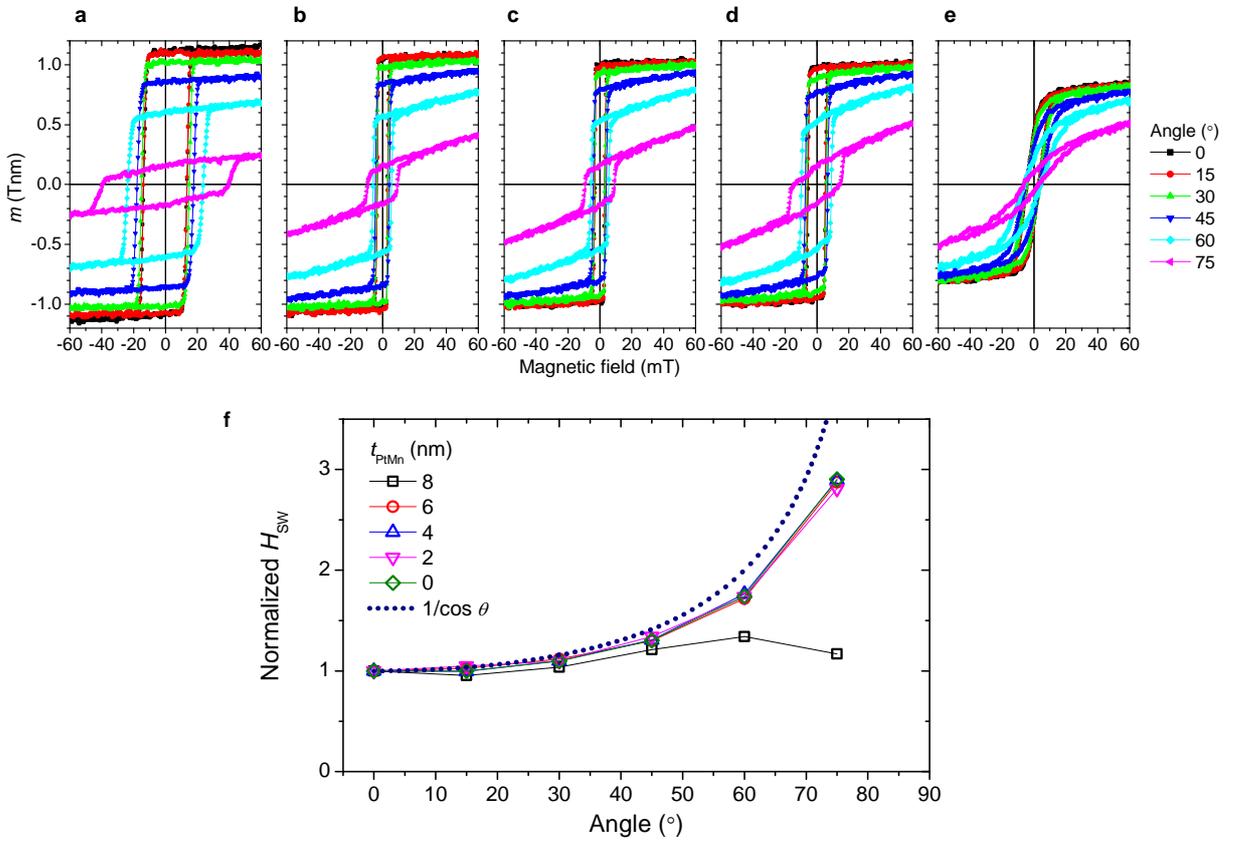

**Figure S6 | Magnetic field angle dependence of *m-H* loops. a-b**, *m-H* loops along various magnetic field directions for $t_{PtMn}$ = 0, 2, 4, 6, 8 nm, respectively. **f**, Switching field $H_{SW}$ as a function of magnetic field angle $\theta$ for various $t_{PtMn}$. Dotted line shows the curve based on Kondorsky model ($H_{SW} \propto 1/\cos\theta$).

**References**

1 Kaplan, B. & Gehring, G. A. The domain structure in ultrathin magnetic films. *J. Magn. Magn. Mater.* **128**, 111, doi:10.1016/0304-8853(93)90863-W (1993).

2 Zhang, C. *et al.* Magnetization reversal induced by in-plane current in Ta/CoFeB/MgO structures with perpendicular magnetic easy axis. *J. Appl. Phys.* **115**, 17C714, doi:10.1063/1.4863260 (2014).

3 Kondorsky, E. On hysteresis in ferromagnetics. *Journal of Physics-USSR* **2**, 161 (1940).